\documentclass[aip, cha,reprint,amsmath,amsfonts]{revtex4-1}

\usepackage{amssymb}
\usepackage{amsmath}
\usepackage{bm}
\usepackage{paralist}
\usepackage{graphicx}
\usepackage{enumerate}
\usepackage{tikz}
\usepackage{graphicx}
\usepackage{verbatim}
\usepackage{etoolbox}

\usetikzlibrary{matrix,positioning,fit}

\usepackage{hyperref}
\hypersetup{
	colorlinks,
	citecolor=blue,
	filecolor=blue,
	linkcolor=blue,
	urlcolor=blue,
	pdfproducer={}
}

\let\bbordermatrix\bordermatrix
\patchcmd{\bbordermatrix}{8.75}{4.75}{}{}
\patchcmd{\bbordermatrix}{\left(}{\left[}{}{}
\patchcmd{\bbordermatrix}{\right)}{\right]}{}{}

\newlength{\Mylen}

\newcommand*{\xdash}[1][3em]{\rule[0.5ex]{#1}{0.55pt}}

\usepackage{ragged2e}
\usepackage{caption}
\usepackage{ragged2e}
\DeclareCaptionJustification{justified}{\justifying}
\usepackage{soul}	

\usepackage [autostyle, english = american]{csquotes}
\MakeOuterQuote{"}

\usepackage[normalem]{ulem}

\usepackage[percent]{overpic}

\usepackage{xcolor}

\usepackage{cprotect}

\begin{document} 
\title{Ensemble-based Topological Entropy Calculation (E-tec)}

\author{Eric Roberts}
\email{eroberts5@ucmerced.edu}

\author{Suzanne Sindi}
\email{ssindi@ucmerced.edu}

\affiliation{School of Natural Sciences, University of California,
	Merced, California, 95343}

\author{Spencer Smith}
\email{smiths@mtholyoke.edu}

\affiliation{Department of Physics, Mount Holyoke College, Massachusetts, 01075}

\author{Kevin A.~Mitchell}
\email{kmitchell@ucmerced.edu}

\affiliation{School of Natural Sciences, University of California,
	Merced, California, 95343}

\date{\today}

\begin{abstract}
Topological entropy measures the number of distinguishable orbits in a dynamical system, thereby quantifying the complexity of chaotic dynamics. One approach to computing topological entropy in a two-dimensional space is to analyze the collective motion of an ensemble of system trajectories taking into account how trajectories "braid" around one another. In this spirit, we introduce the Ensemble-based Topological Entropy Calculation, or E-tec, a method to derive a lower-bound on topological entropy of two-dimensional systems by considering the evolution of a "rubber band" (piece-wise linear curve) wrapped around the data points and evolving with their trajectories.  The topological entropy is bounded below by the exponential growth rate of this band. We use tools from computational geometry to track the evolution of the rubber band as data points strike and deform it. Because we maintain information about the configuration of trajectories with respect to one another, updating the band configuration is performed locally, which allows E-tec to be more computationally efficient than some competing methods. In this work, we validate and illustrate many features of E-tec on a chaotic lid-driven cavity flow. In particular, we demonstrate convergence of E-tec's approximation with respect to both the number of trajectories (ensemble size) and the duration of trajectories in time.
\end{abstract}

\maketitle

\noindent \textbf{From the stirring of dye in viscous fluids to the availability of essential nutrients spreading over the surface of a pond, nature is rife with examples of mixing in two-dimensional fluids. The long-time exponential growth rate of a thin filament of dye stretched by the fluid is a well-known proxy for the quality of mixing in two dimensions. In the real-world study of mixing, this stretching rate may be hard to compute; the velocity field may not be known or may be expensive to recover or approximate, thus limiting our knowledge of the governing system and underlying mechanics driving the mixing. One alternative is to use time-ordered trajectory data, often obtained from tracer particles such as ocean drifters.  In this paper, we use the collective motion of such trajectories, along with tools from computational geometry, to develop a lower bound to the stretching rate.  The lower bound is obtained by approximating the filament of dye with a piece-wise linear, non-intersecting "rubber band" stretched around the data points. We call our algorithm the Ensemble-Based Topological Entropy Calculation, or E-tec.}
\begin{center}
\xdash[22em]
\end{center}

\section{Introduction}
A variety of techniques have been used to quantify complexity and uncertainty in dynamical systems theory. These tools include the finite-time Lyapunov exponent (FTLE) field\cite{ding2007nonlinear, eckmann1995fundamental}, which measures the exponential rate of separation between points in a small neighborhood; the finite-time entropy (FTE) field\cite{froyland2012finite}, a probabilistic approach to measuring local stretching and determining the uncertainty in a trajectory's final position; operator-theoretic methods, such as the eigenfunctions and eigenvalues of the Koopman operator\cite{arbabi2016ergodic}; and numerical evolution of a two-dimensional material-curve, who's growth rate is shown to be equivalent to the topological entropy \cite{candelaresi2017quantifying,yomdin1987volume,newhouse1988entropy, newhouse1993estimation, newhouse1989continuity}, which measures the proliferation of distinguishable orbits\cite{bowen1971periodic}. Such knowledge aids greatly in a wide variety of natural and industrial fluid systems, including the large-scale dispersion of pollutants in the Earth's atmosphere and oceans \cite{aref2017frontiers}; for example, understanding how regions of fluid remain isolated from each other helps predict the fate of oil spills \cite{olascoaga2012forecasting, froyland2012three}. Understanding mixing in the rapidly developing field of microfluidics \cite{lee2011microfluidic, sanchez2012self} could lead to new classes of self-mixing active solvents that further our understanding of the kinetics of mass transport and chemical reactions. Obvious industrial applications include the optimization of stirring devices in viscous fluids, such as the rod-stirring devices used to effectively knead dough, pull taffy \cite{finn2011topological, thiffeault2018mathematics}, or manufacture glass compounds\cite{finn2011topological, chu2018topological}.

However, a problem remains for many techniques --- the fine-scale structure of a system may not appear without a high point density. A sufficient number of system trajectories and/or the linearizations about these trajectories may simply be too expensive to compute or to measure experimentally. We seek techniques that can accommodate such sparse data. 

Our goal is to compute material-line stretching rates using only 2D particle trajectories, like those collected from oceanic floats\cite{ledwell1993evidence, thiffeault2010braids} or fluorescent beads in microfluidic systems\cite{kelley2011separating, ouellette2008dynamic}. These data sets may be sparse, and hence may not fully sample all of the 2D space. We are motivated by Budi\v{s}i\'{c}, Allshouse, and Thiffeault \cite{budivsic2015finite, thiffeault2010braids, allshouse2012detecting}, who use braiding theory to compute a lower bound for topological entropy of flows from such data sets. The initially embedded material-curve is thought of as an elastic line whose growth rate is computed using the collective motion of all available trajectories moving through space in concert. In essence, the relative motion of an ensemble of trajectories in space encodes global information that is not contained in any one individual trajectory. That is, extra information is "hiding" in an ensemble of trajectories, which is not exploited in a trajectory-by-trajectory approach. 

In this paper, we focus on these underlying stretching and folding processes that drive mixing in two dimensional fluids. We apply computational geometry techniques to develop a 2D algorithm titled the Ensemble-based Topological Entropy Calculation (E-tec), which may be downloaded at \href{https://zenodo.org/badge/latestdoi/146612307}{10.5281/zenodo.1405656}.  E-tec achieves three main goals: a) estimation of a lower bound to the topological entropy on data sets, ~ b) convergence to the topological entropy as ensemble size increases, ~c) linear scaling in runtime with the length of trajectories and $N^k\log{N}$ scaling with the number of trajectories $N$. (Values of $k$ range from $1/3 \leq k \leq 3/2$ and typically $k \lesssim 1$. We point the reader to Appendix~\ref{sect:appendix} for a discussion.) E-tec does not require the flow to be area preserving or incompressible.

The remainder of this paper is broken up into six sections. We first review topological entropy (Sect.~\ref{sect:topologicalentropy}) and then summarize (Sect.~\ref{sect:etecoverview})  and give procedural details (Sect.~\ref{sect:etecsteps}) of our E-tec algorithm. We next evaluate the performance of E-tec on a chaotic, lid-driven cavity flow as a test case (Sect.~\ref{sect:results}) and show that results are consistent with the braiding approach. Finally, we demonstrate E-tec's robustness and show evidence that the E-tec runtime compares favorably to braiding algorithms (Sect.~\ref{sect:discussion}). Appendix~\ref{sect:appendix} contains details regarding E-tec's runtime scaling and computational bottlenecks.

\section{Topological Entropy} \label{sect:topologicalentropy}

Topological entropy is a measure of the growth rate of the number of distinguishable orbits\cite{young2003entropy}.  More formally, topological entropy is defined by considering equivalence classes of trajectories of duration $T$ that are only distinguished if they are, at \emph{any} point in time, further than some resolution $\epsilon>0$ apart. The number of these $\epsilon$-distinct classes of trajectories increases as both $T \rightarrow \infty$ and $\epsilon \rightarrow 0$. Topological entropy measures the growth of all $\epsilon$-distinct trajectories as $T \rightarrow \infty$. Specifically, the topological entropy $h$ is the exponential growth rate in time of the number of distinct trajectory classes for arbitrarily small $\epsilon$.

In two-dimensional flows, topological entropy $h$ can be estimated by embedding an initial material-curve, e.g. a line of dye, of length $L_0$ in the system and estimating its growth under the evolution of the flow \cite{newhouse1993estimation}. At long times, the length $L(t)$ of the curve as a function of time $t$ grows exponentially as
			\begin{equation} \label{h_t}
			L(t) ~ \approx ~ L_0~e^{ht}.
			\end{equation} 
Thus, direct computation of the curve's evolution is troublesome in chaotic flows since the length is expected to grow exponentially fast, which requires an exponentially growing number of trajectories to maintain sufficient point density of the curve. Other techniques for extracting topological entropy operate on a trajectory-by-trajectory basis, i.e. ensemble-averaging some quantity (such as the Jacobian singular values) computed one trajectory at a time.  This is the approach taken in recent work on {\it expansion entropy\cite{hunt2015defining} }, a generalization of topological entropy, which unlike Eq.~\ref{h_t}, scales to higher dimensions for all flows and requires no computing or measuring of multidimensional surfaces.

As an alternative approach for 2D systems, a lower bound to the topological entropy may be computed with a finite number of trajectories and no detailed knowledge of the velocity field. The material-curve to be advected is represented by a taut elastic loop that wraps tightly around trajectories that strike it. Since an advected material-curve may be continuously deformed into this taut loop given the same trajectory evolution, the need for maintaining material-curve point density is eliminated. The loop is stretched and folded over itself exponentially many times in a chaotic flow. Its exponential growth rate is a lower bound to the full system's topological entropy\cite{thiffeault2010braids}. 

In this more topological setting, braiding theory has been used to compute this lower bound. The Finite-Time-Braiding-Exponent (FTBE) method \cite{budivsic2015finite} evolves the loop forward using the entanglement of a finite number of trajectories. However, this method scales quadratically in the number of points $N$ due to the braid approach requiring $\mathcal{O}(N^2)$ algebraic generators per unit time. This renders braiding exponent calculations unwieldy for systems requiring many trajectories. 

\begin{figure*}[]
	\centering
	\includegraphics[width=1\linewidth]{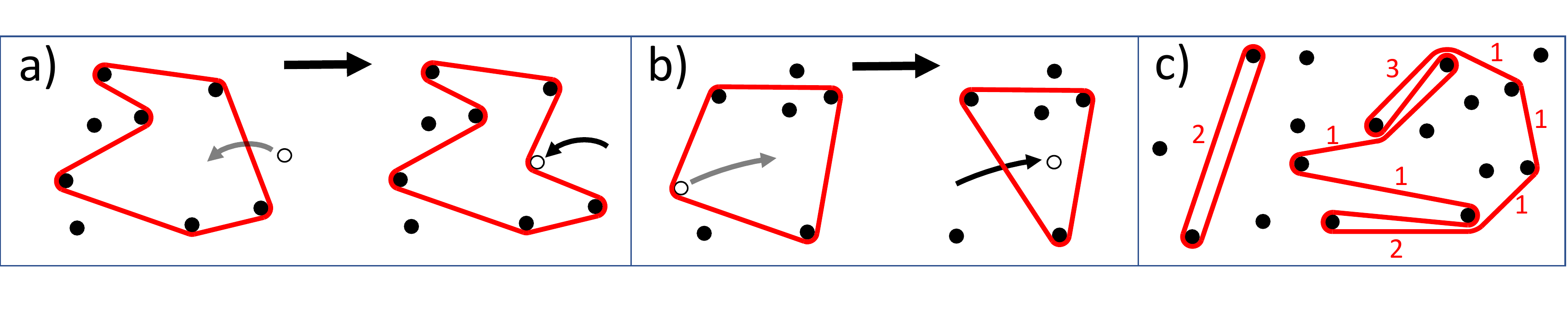}

	\caption{\textbf{Band Deformation.} a) The white point strikes and deforms the band (red). ~b) The white point detaches from the band. Notice the band edge is taut after detachment. ~c) An initial rubber band stretched between two points on the left with edge weights displayed. A more complicated band on the right. The edge weights correspond to the number of times the band crosses an edge. }
	\label{fig:InitialBands}
\end{figure*}

To develop a computationally efficient method to estimate a lower bound on the topological entropy of a planar flow that scales sub-quadratically in the number of points $N$, we compute the stretching rate of an advected elastic curve directly. Referring now to the elastic curve or loop as a rubber band, we use the same FTBE idea of trajectories working in concert to stretch and fold the band. The E-tec algorithm achieves this using the same input: i) a set of (typically aperiodic) trajectories $\{x_i(t), y_i(t)\}$ that are discretized over time $t_1, t_2,\dots$ and ~ii) a user-specified, non-self-intersecting elastic band which wraps around a set of trajectories. The output is the number of edge segments in the band as a function of time. However, instead of using a braid representation to compute the stretching of the band, E-tec computes this stretching, and thus the topological entropy, \emph{directly} by using a triangulation to detect all point-band collisions. 

In summary, E-tec tracks the crossing of a trajectory with \emph{only} its neighboring edges in the triangulation, unlike the braiding method which concerns itself with \emph{each} trajectory's relative position with every \emph{other} trajectory along a projection axis. This idea leads to a more favorable sub-quadratic runtime scaling of $\mathcal{O}(N^k \log N)$, where $1/3 \leq k \leq 3/2$. (For a detailed discussion about the two methods' runtime scaling in the number of points, we refer the reader to Appendix~\ref{sect:appendix}.) The idea of using an advected dynamic triangulation to compute topological entropy was first proposed by Marc Lefranc\cite{lefranc2013reflections, lefranc2006alternative, lefranc2008topological}. Lefranc's work was restricted to the entropy generated by periodic orbits, and he did not develop a general algorithm to implement this. To our knowledge, this work is the first attempt to fully generalize Lefranc's ideas to aperiodic orbits.

\section{Overview of E-tec} \label{sect:etecoverview}

\begin{figure}[t]
	\centering
	
	\begin{overpic}[width=0.48\textwidth]{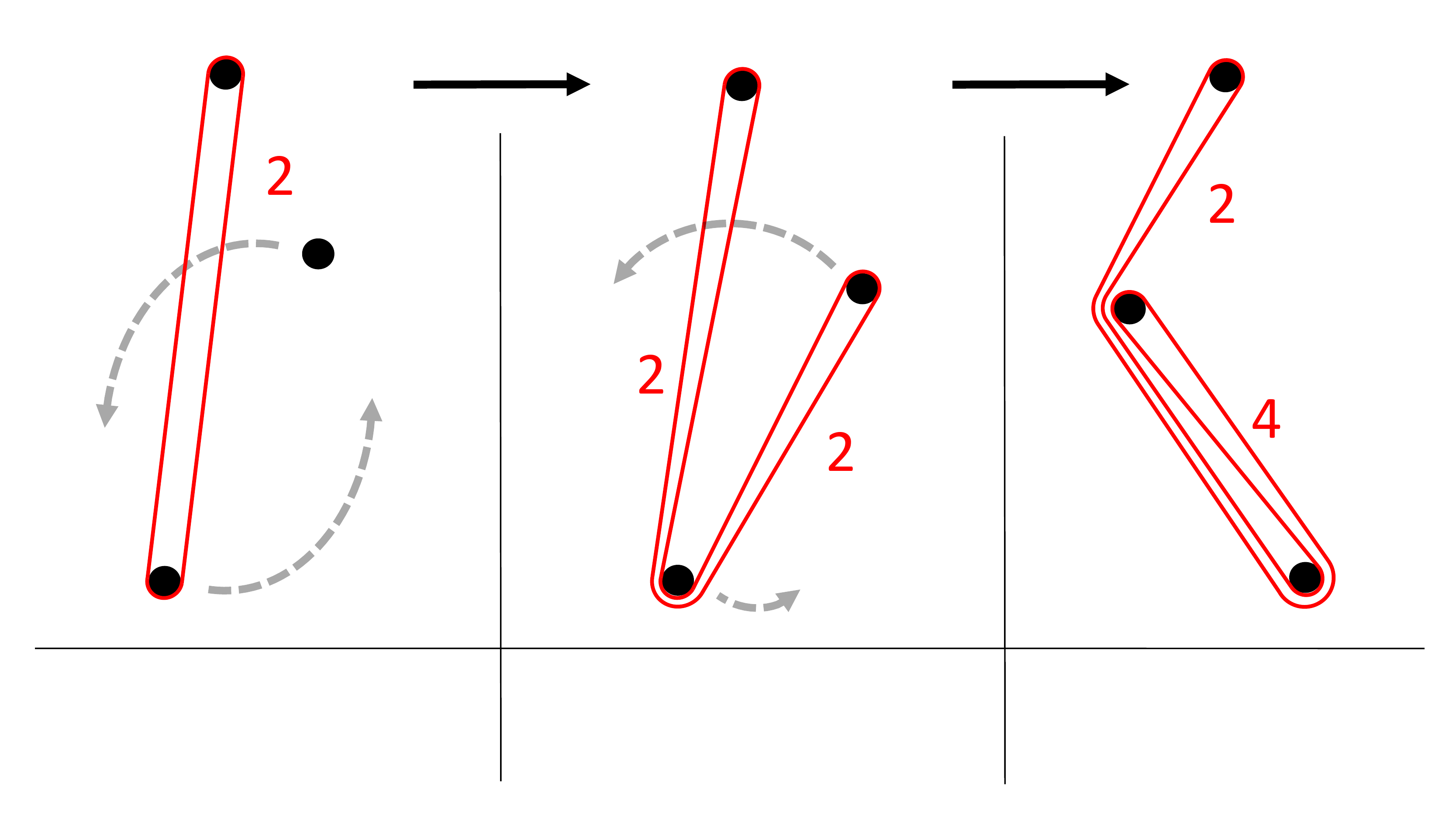}
		\put (3, 7) {\begin{small}Two total edges\end{small}}
		\put (38,7) {\begin{small}Four total edges\end{small}}		
		\put (73,7) {\begin{small}Six total edges\end{small}}
	\end{overpic}

	\caption{\textbf{Edge Weights.} E-tec counts the number of edges of a rubber band as it is stretched by moving points. As the two bottom points rotate, the red band, initially wrapped around two points, is stretched and folded (left to right). E-tec tracks the growth of this band by assigning a weight to each edge corresponding to the number of times the band passes over this edge.}
	\label{fig:bandsweight}
\end{figure}

We first give an overview of E-tec and forgo the details to the next Section (Sect.~\ref{sect:etecsteps}). E-tec computes how an initial, closed, piecewise linear, non-self-intersecting rubber band in $\mathbb{R}^2$ evolves under an ensemble of trajectories. The vertices of the band coincide with trajectories from the ensemble. When trajectories strike the band, they do not penetrate it but stretch it like a piece of elastic (Fig.~\ref{fig:InitialBands}a). In this manner, the band is stretched and folded, typically producing a growing number of edges wrapping around each other. Our algorithm tracks the configuration of the band. Care must also be taken in finding when and how a trajectory detaches from an edge.  This detachment results in two band edges returning taut (Fig.~\ref{fig:InitialBands}b), in much the same way a tight string will return taut once plucked (stretched) and released (undoing the stretching). Each band edge is assigned an integer weight $\omega$ indicating the number of times the band stretches across it (Fig.~\ref{fig:InitialBands}c and Fig.~\ref{fig:bandsweight}). For chaotic advection, the total weight of the band will grow exponentially, as shown in Sect.~\ref{sect:results}. This exponential growth rate is a lower bound to the true topological entropy of the dynamical system. Even though the weight of all the edges grows exponentially, the number of unique edges is bounded. 

\begin{figure}[t]
	\centering
	\begin{overpic}[width=.95\linewidth]{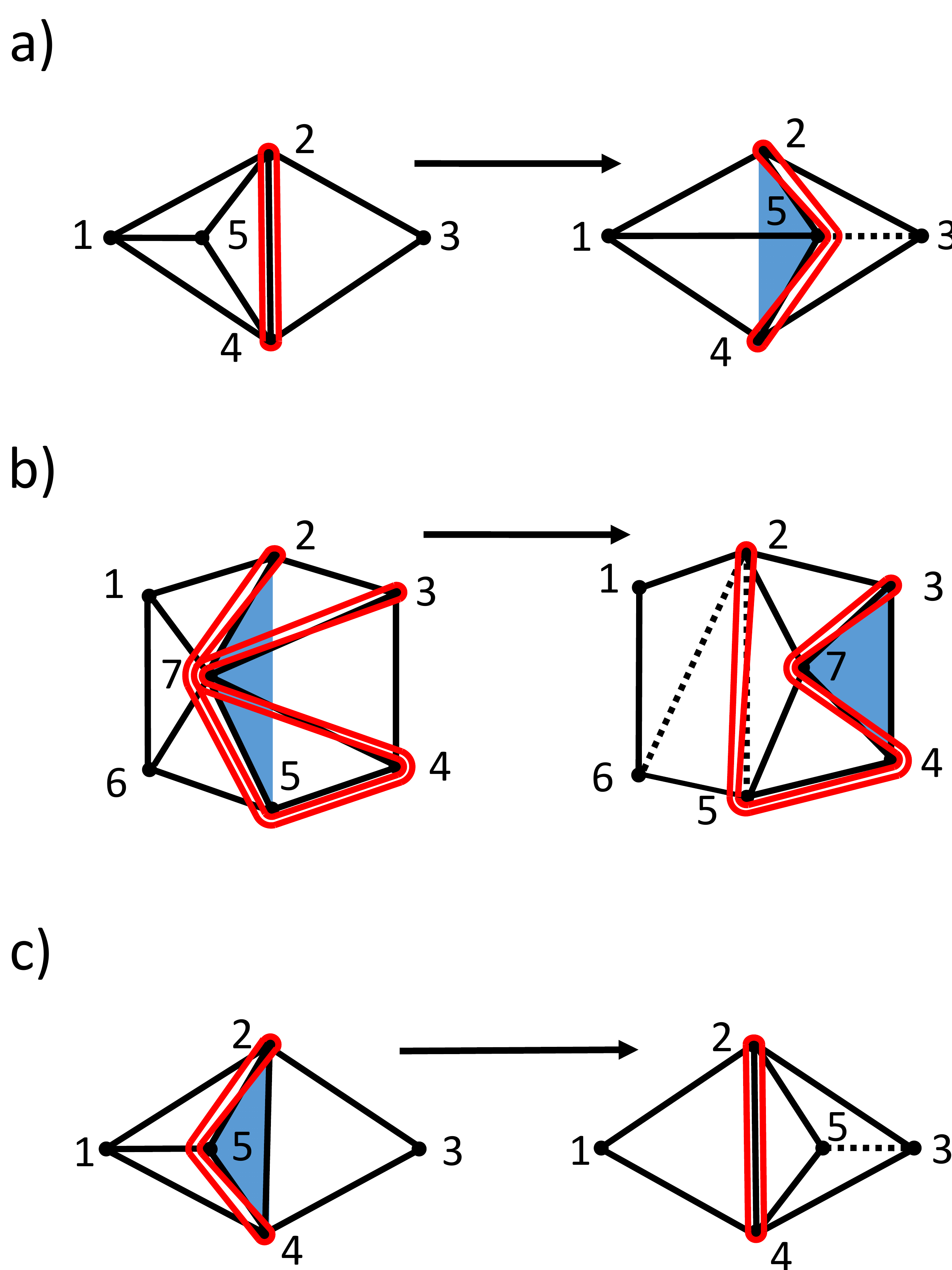}
		\put (29.5, 95) {\begin{normalsize}\textbf{Core Triangle}\end{normalsize}}
		\put (29, 63) {\begin{normalsize}\textbf{Outer Triangle}\end{normalsize}}
		\put (11, 24) {\begin{normalsize}\textbf{Combined Core and Outer Triangle}\end{normalsize}}
	\end{overpic}
	\caption{\textbf{Events of E-tec Algorithm.} a) As point 5 moves right, triangle (2, 4, 5) collapses and inverts orientation. Two core triangles are re-triangulated, with the new edge shown as dashed.  The initial edge weight of 2 for segment (2, 4) is shifted to segments (2, 5) and (4, 5). The blue-highlighted triangle (2, 4, 5) is the new outer triangle of point 5.  It records which triangle collapse would be needed for the band to "snap back" taut, thereby undoing the collision. ~b) As point 7 moves to the right, outer triangle (2,5,7) collapses and the band edges (2, 7) and (5, 7) straighten into (2, 5). The three core triangles within pentagon (1,2,7,5,6) are reconfigured into three new core triangles (1,2,6), (2,5,6), and (2,5,7). Point 7 is still a candidate for future detachment, with new outer triangle (3,4,7), which also happens to be a core triangle. ~c) In blue is a combined core and outer triangle (2,4,5). As point 5 moves to the right and this triangle collapses, the band returns taut around segment (2,4). Three core triangles (1,2,4), (2,3,5), and (3,4,5) are reconfigured, with the new edge shown as dashed. Collapsed triangle (2,4,5) (previously shaded) remains as a core triangle.}
	\label{fig:Events}
\end{figure}

\begin{figure}[t]
	\centering

	\includegraphics[width=1.05\linewidth]{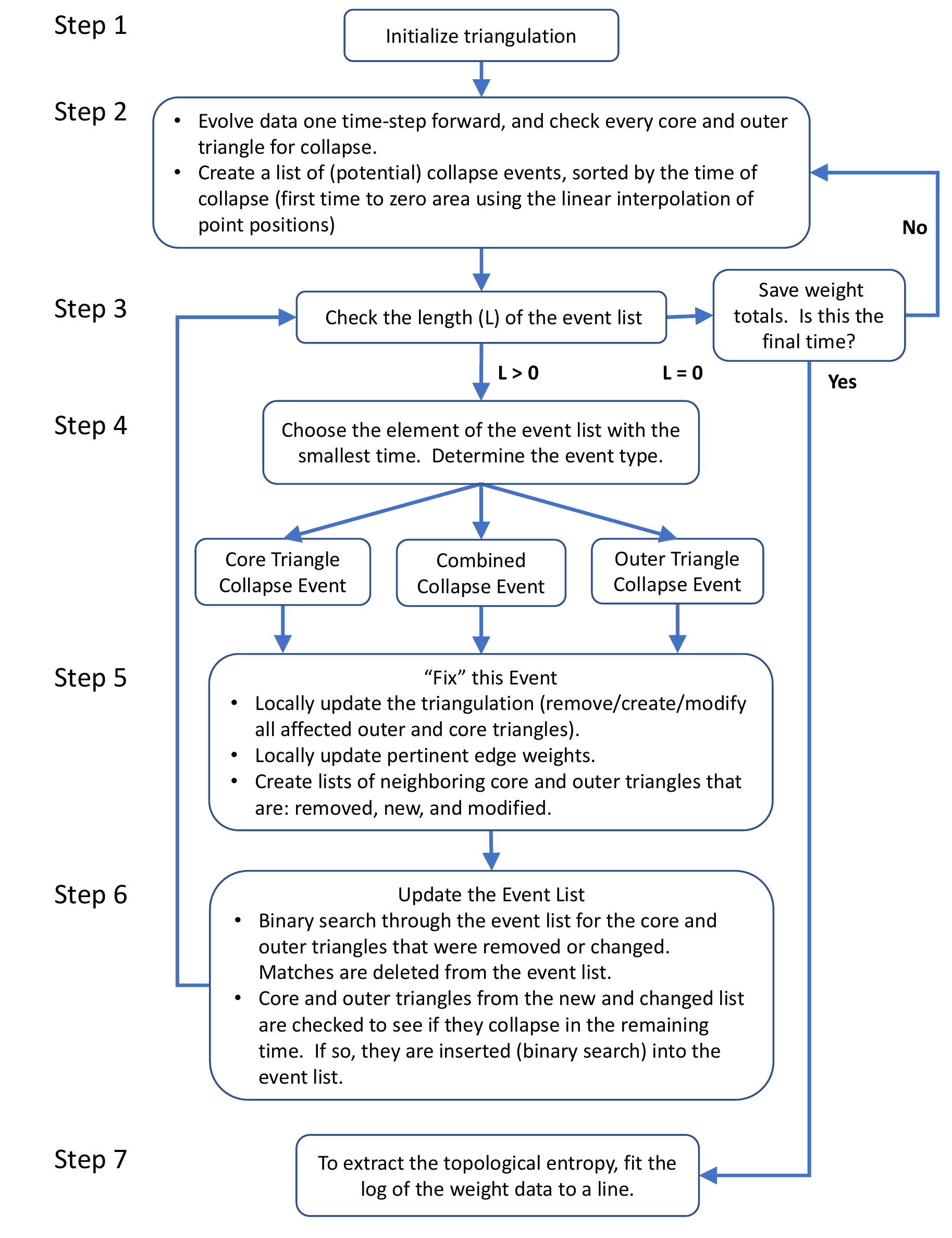}
	\caption{ \textbf{E-tec Algorithm Flowchart.} As described in Sec.~\ref{sect:etecoverview}, E-tec employs computational geometry techniques for tracking the evolution of a piecewise-linear band. Full details are given in Sect. \ref{sect:etecsteps}.}
	\label{fig:Flowchart}
\end{figure}
	
 E-tec efficiently tracks band growth by simply shifting edge weights to the appropriate edges when a point collides with, or detaches from, the band. A key component of the algorithm is the detection of all relevant point-edge collisions. We achieve this by maintaining a triangulation of all trajectories for all times. First, edge weights are determined corresponding to the initial placement of the band. Next, the data points may be triangulated in any manner consistent with the initial placement of the band. For any initial band, E-tec's computation of the evolved band is independent of the initial constrained triangulation. Here, we initialize with a constrained Delaunay triangulation\cite{toth2017handbook}. Note that  the algorithm is fast enough to run and compare many different initial bands in a reasonable time.
 
 The triangles that make up the triangulation are called \emph{core} triangles. Each edge of the stretched band lies within the triangulation, so that each time a point strikes the band, the orientation of one of the core triangles will be inverted. We refer to this inversion as a triangle collapse. All band deformations will be detected since band edges remain in the core triangulation. The triangulation must be updated upon any triangle collapse. This update is \emph{local} to the detection of each event, resulting in the rearrangement of edges and triangles near the collision only (illustrated in Fig.~\ref{fig:Events}a). Similarly, the only edge weights that are shifted are those involved in the collision. The update process is independent of both the number of points $N$ and the number of triangles. 
	
In addition to collisions, we need to detect when a trajectory detaches itself from a band edge. E-tec records which edges of the band are candidates for detachment by storing the triangle made up of the outer-most band edges attached to each point, i.e. typically the most recent edges to have struck a point. These triangles are called \emph{outer} triangles and are shown in blue in Fig.~\ref{fig:Events}. Unlike the core triangles, the outer triangles do not form a triangulation of space. Rather, there is simply one outer triangle for each vertex crossed by the band. When a point detaches from the band, its corresponding outer triangle collapses and inverts its orientation. After the outer band edge peels off the point, there may remain other band edges still wrapped around the point. (Follow point 7 in Fig.~\ref{fig:Events}b for an example.)  In this case, E-tec recalculates and stores the new outer triangle.  Note that the outer triangle of a given point can always be recalculated from just the weights of all edges adjacent to the point.  Thus, E-tec must track when both core and outer triangles collapse.

The triangulation update process following an outer triangle inversion remains local, though the process differs from the core triangle inversion update in one fundamental aspect: the local re-triangulation is constrained to contain the band that remains taut. This creates a possibly non-unique choice in edges needed to complete the triangulation. As an example, notice that edge (1, 5) could have replaced edge (2, 6) to complete the triangulation in Fig.~\ref{fig:Events}b. Because of this, E-tec will not generally recover the initial triangulation away from the band if trajectories are run forward and then exactly backward in time. However, the algorithm \emph{is} time-reversible in that the band returns to its initial configuration after running the trajectories backwards to their initial positions.

In summary, there are two kinds of events that must be detected: the collapse of either a core or outer triangle. In the given time interval, these events are detected by finding the time for which their area first goes through zero. This time of first collapse is simply the appropriate root of the area quadratic polynomial, which is formed from the linear interpolation of triangle point positions. (For any reader interested in the scaling of the number of events with the number of trajectories used, we refer them to Appendix~\ref{sect:appendix}.) Once these events are detected, they are put in a time-sorted list and processed in order. Each event is "fixed" by locally updating the core triangulation, outer triangles, and edge weights. In the course of fixing an event, we may need to add or remove events from the event list. Event lists become large for densely-packed ensembles, though E-tec parses through each event and performs each subsequent triangulation update efficiently, as verified in Sect.~\ref{sect:results}. A flowchart summarizing the E-tec algorithm is given in Fig.~\ref{fig:Flowchart}. The algorithm steps found here are detailed in the following section.
		
\section{E-tec Algorithm Details} \label{sect:etecsteps}

This section details our implementation of the E-tec
algorithm. \\
%
\noindent \textbf{Input:} The following inputs are required by the algorithm:

\begin{compactenum}
	\item The precomputed (or experimentally measured) trajectories.
	
	\item An initial, non-self-intersecting rubber band stretched around a sequence of data points, specified by the set of edges connecting pairs of data points. This is represented as a counterclockwise ordering of this set of points. It is often convenient to choose an initial band that encloses two distant points.
	
\end{compactenum}

\noindent \textbf{Output:} E-tec tracks the evolution of the band, as we will describe below, and outputs:

\begin{compactenum}
	\item The state of the stretched rubber band as a function of time, recorded as a (core) triangulation of all data points and a set of edge weights of this triangulation.
	
	\item The sum of all band edge weights $\omega$ as a
	function of time.
	
	\item The exponential growth rate of the band (topological entropy),
	determined by the slope of the best fit line for the $\ln(\omega)$ vs. time graph.
	
\end{compactenum}

\noindent \textbf{Data structures:} E-tec maintains the following data
structures as a function of time:

\begin{compactenum}
	\item A \emph{core} triangulation of all data points in the plane.
	
	\item The weights on each edge in the triangulation.  (Non-zero weighted edges constitute the stretched rubber band.)
	
	\item For each relevant data point, the \emph{outer band} triangle (abbreviated
	\emph{outer} triangle) records the outermost wrapping of the rubber band around that point. (See the blue shaded triangles in Fig.~\ref{fig:Events}.)  During the algorithm's run, the outer triangle represents the piece of rubber band that has struck the point most recently and hence is a candidate for detachment at a future time. For example, upon inspection of vertex 7 in Fig. \ref{fig:Events}b, we may deduce that of all the red band edges attached to it, the two that created the largest angle would be the ones to snap back and revert to a single edge. Specifically, edge $(2,5)$ will snap back taut if triangle $(2,5,7)$ changes orientation. Notice that outer triangles are not necessarily contained in the set of all core triangles.
	
\end{compactenum}

\noindent \textbf{Steps:} We outline the key steps taken by E-tec in tracking the evolution of a rubber band. These steps are summarized in the Fig.~\ref{fig:Flowchart} flowchart.

\begin{figure}[t]
	
	\centering
	\includegraphics[width=.85\linewidth]{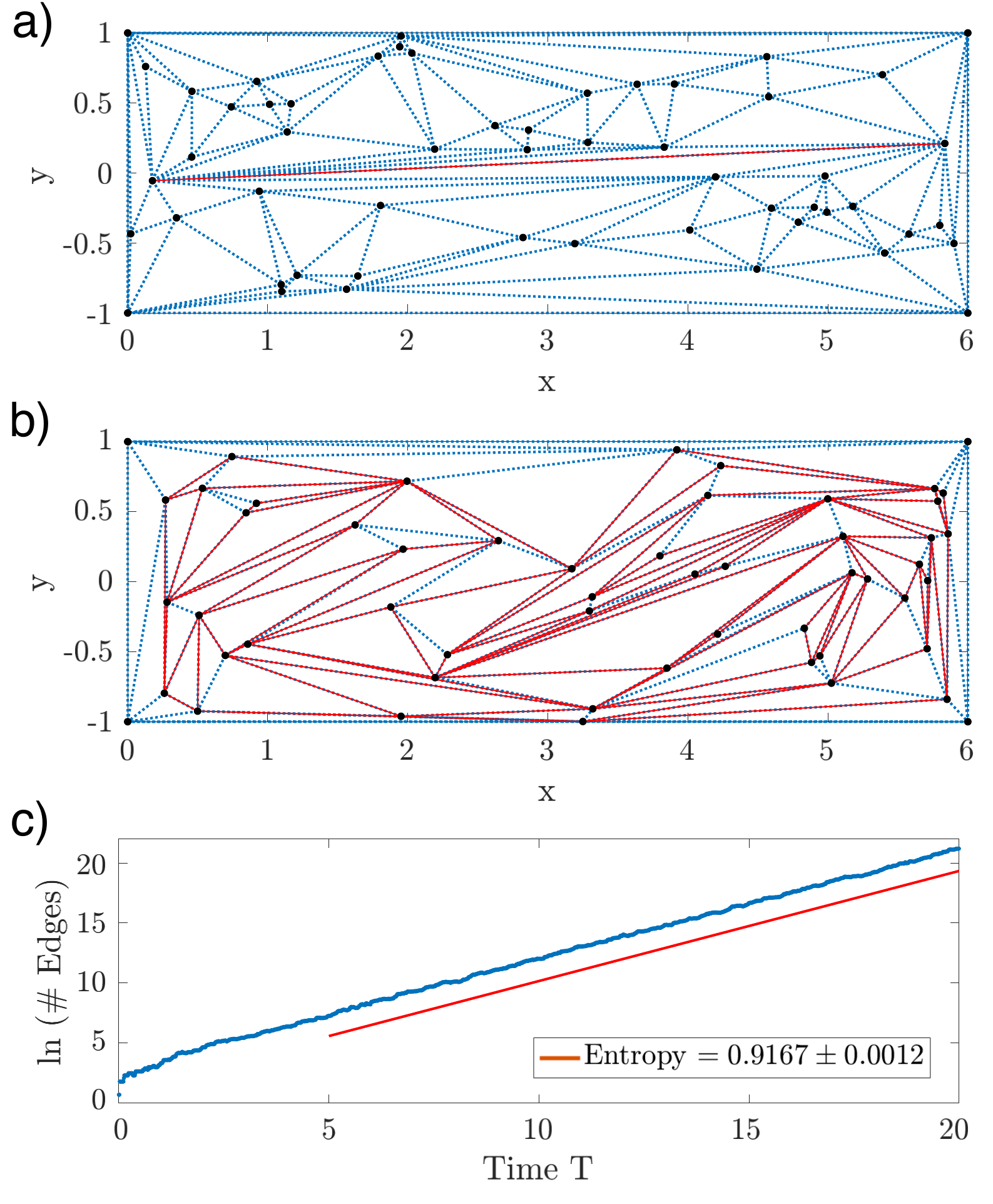}
	\caption{\textbf{Numerical Example of an E-tec Implementation.} a) Initial data points with the band wrapped around two points (in red). The core Delaunay triangulation (in blue, dotted) is constrained to include the red band edge.  ~b) Final data point positions at $T = 20$, the triangulation, and the stretched band evolved under the motion of the trajectories. Dynamics is given by model in Sect.~\ref{sect:results} with $\tau_f = 0.96$. ~c) E-tec output: the number of band edges as a function of time (blue). The slope of the best-fit line (red, dashed) is the topological entropy estimate. Please see online multimedia view for the related movie.}
	\label{fig:Example}
\end{figure}

\begin{compactenum}
	\item We first initialize the core triangulation using a constrained Delaunay triangulation \cite{toth2017handbook} of the initial points (with the initial placement of the rubber band as the constraint). See Fig.~\ref{fig:Example}a.
\end{compactenum}

\noindent In steps (2-6) we evolve the state of the system (core triangulation, weights, and outer triangles) forward using the next time-slice in the trajectory data as input.  Notice that E-tec does not need the whole trajectory at once in order to evolve the triangulation forward, and therefore could be used in real-time during experimental data collection.

\begin{compactenum}	
	\item[2.] For each core and outer triangle in the current state of the system, we use the linear interpolation of point positions to determine if and when a triangle will pass through zero area during this time step. These collapse events are sorted by time into an event list.
	
	\item[3.]  If the event list is non-empty, we go to step 4 and determine the event type of the next collapse event. If the event list is empty, we then add up the weights of every edge to get the current total weight $\omega$ of the band, and store this value. This acts as a proxy for the length of the band, and grows with the same exponential rate in time. If we are at the final time of the trajectory data, we end by analyzing the accumulated weight data in step 7. Otherwise, we move on to the next trajectory time in step 2.
	
	\item[4.] A collapse event can be one of three general types: a core triangle collapse (Fig.~\ref{fig:Events}a), an outer triangle collapse (Fig.~\ref{fig:Events}b), or a combined core and outer triangle collapse (see Fig.~\ref{fig:Events}c for an illustration).  While the specifics of how the three types of collapse events are handled are different, the broad strokes, as seen in step 5, are the same. 
	
	\item[5.] For each collapse event type, there is a general template for adding, removing, and/or modifying the core and outer triangles that are adjacent to the collapsing triangle.  Crucially, this process is local, and the number of operations is bounded and does not grow with the number of trajectories.
	
	\item[6.] The local deletions, creations, and modifications of core and outer triangles that result from handling a collapse event potentially affect the overall event list for this time-step.  First we consider the deleted and modified core and outer triangles.  If, before modification, they have a time-to-zero-area that is in the remaining fraction of the current time-step, then we search for and remove them from the event list.  Next we consider the new and modified core and outer triangles. If, after modification, they will collapse in the remaining time-step, we search for the proper position to insert them into the sorted event list.  Both searches are binary, and constitute one of the two aspects of the algorithm that give us $\mathcal{O}(N^k\log N)$ computational complexity, where $1/3 \leq k \leq 3/2$  ($\mathcal{O}(\log N)$ for binary search and $\mathcal{O}(N^k)$ searches per time-step).  After modifying the event list, we return to step 3.
	
	\item[7.] Approximate the topological entropy by computing the exponential growth rate for the total weight over time.
	
\end{compactenum}

\section{E-tec Algorithm Verification}

\label{sect:results}

\begin{figure}[t]
	\centering
	\includegraphics[width=1\linewidth]{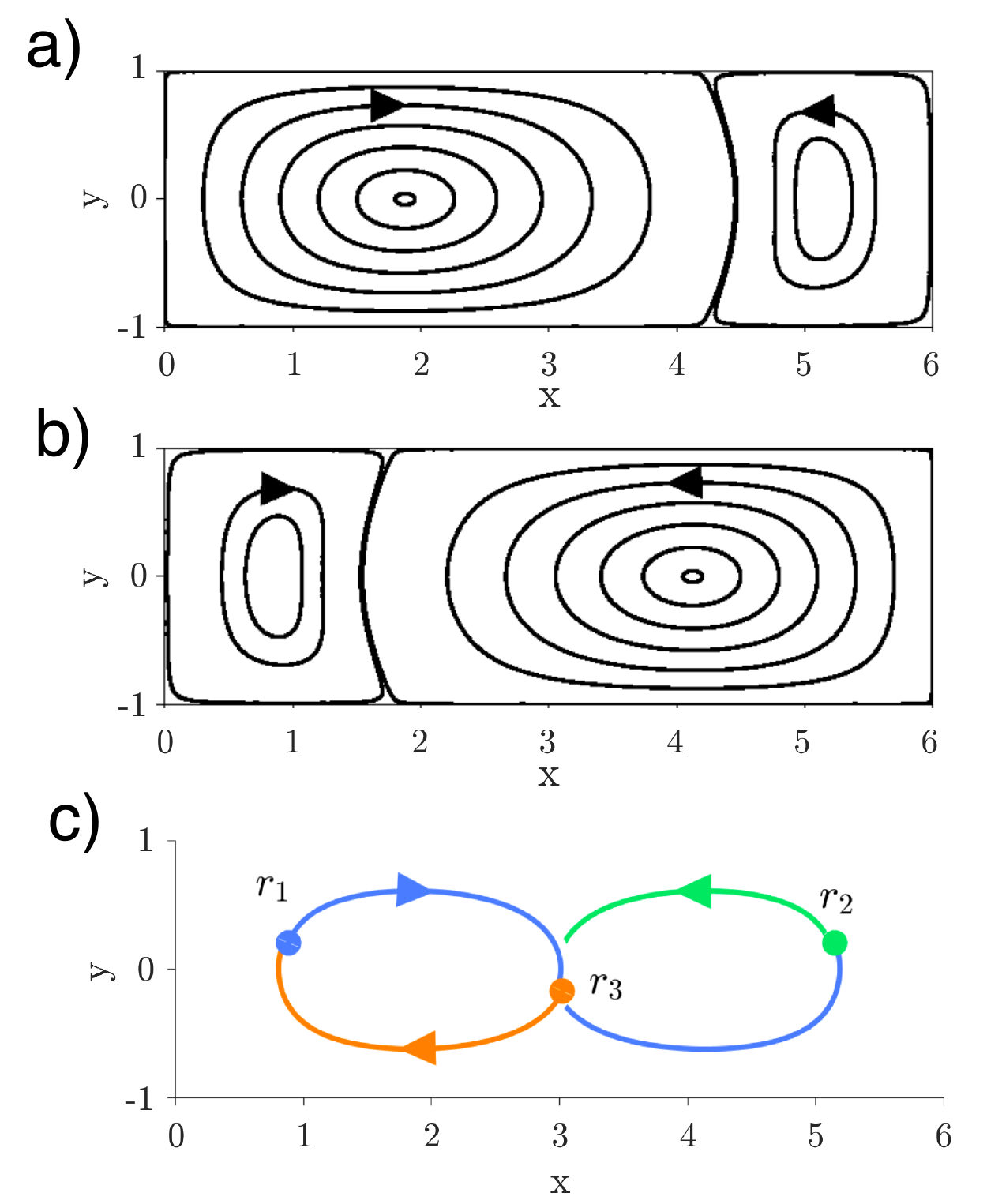}
	
	\caption{\textbf{Dynamics of Chaotic Lid-Driven Cavity Flow.} We depict streamlines of the flow, Eq. (\ref{stremler}). a) Motion under the first half-period, $n\tau_f \leq t <(n + 1/2)\tau_f $.  b) Motion under the second half-period, $(n + 1/2)\tau_f \leq t <(n+1)\tau_f$.~ c) Illustration of a period-three orbit $r_i$.  Each color (blue, green orange) represents the trajectory evolving forward one period. }
	\label{fig:streamines}
\end{figure}
 
 \begin{figure*}[t]
 	\centering
 	\includegraphics[width=1\linewidth]{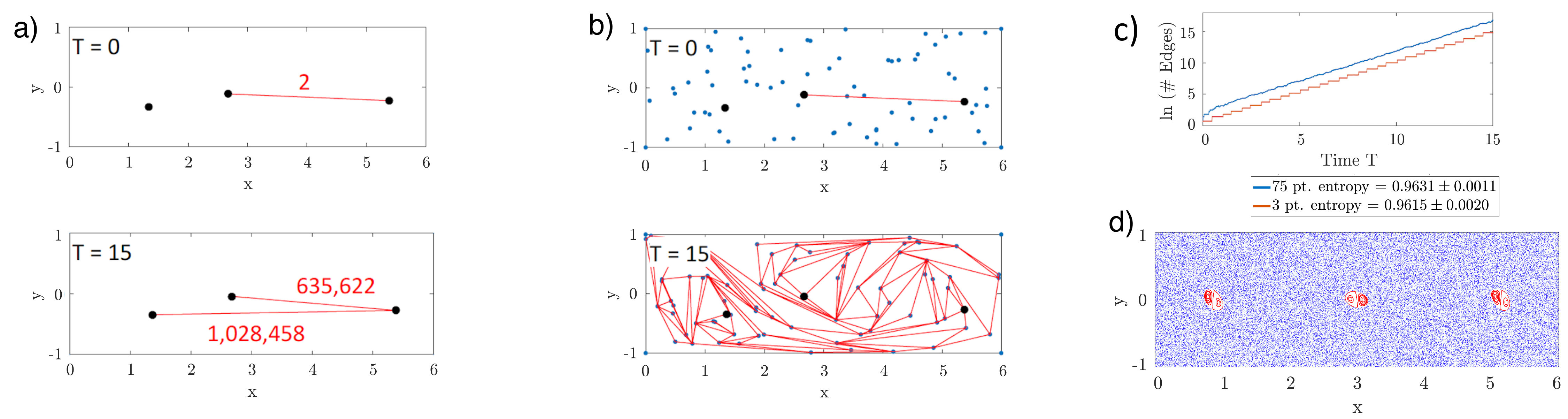}
 	\caption{\textbf{E-tec Analysis of the Chaotic Lid-Driven Cavity Flow.} We show E-tec results on trajectories governed by Eq.~(\ref{stremler}) with $\tau_f = 0.96$, guaranteeing the existence of a period-three orbit, seen in Fig~\ref{fig:streamines}c. a) We show E-tec results when considering only 3 points close to the period-three orbit and contained in period-three islands. We consider an initial band around the two right points (top). This band evolves (bottom) into a highly stretched band (edge weights in red) around all three points by $T = 15$. ~b) We consider the same 3 initial points, but add 72 random trajectories (top). The dynamics are more complex (bottom, weights omitted). ~ c) The growth rate in the number of edges, i.e. our estimate of the topological entropy, for (a, red) and (b, blue) is the same. This indicates the entropy is driven by the period-three islands as also shown by Ref. \citenum{sattari2016using}. ~d) The coherent period-three islands, noted in Refs. \citenum{sattari2016using}, \citenum{grover2012topological}, and \citenum{stremler2011topological}, are clearly seen in the Poincar\'{e} return map of a long-lived trajectory in blue. Ten long-live trajectories inside the islands are shown in red.}
 	\label{fig:etecper3}
 \end{figure*}
 
 In this section, we verify the E-tec algorithm by running E-tec on numerical trajectories sampled from a chaotic lid-driven cavity flow used to study chaotic advection \cite{grover2012topological}.  A numerical example of E-tec applied to real trajectory data (requiring only seconds to run) is shown converging to the theoretical topological entropy lower bound of the flow in Fig.~\ref{fig:Example}. In later subsections, we compare our results to lower bounds on topological entropy computed from two different methods; first, by a direct application of Eq.~(\ref{h_t}) to a growing material-line, and second, by a technique called homotopic lobe dynamics (HLD), which extracts symbolic dynamics from finite-length pieces of stable and unstable manifolds attached to fixed points of the fluid flow\cite{sattari2016using, mitchell2009topology, mitchell2012partitioning}. 

\subsection{Chaotic Lid-Driven Cavity Flow}  \label{sect:stremler}

The chaotic lid-driven cavity model\cite{grover2012topological, stremler2011topological, rao2012mixing,meleshko2004infinite} is a two-dimensional area-preserving flow defined over a 2D vertical cross-section of a rectangular cavity, extending vertically from $-b \leq y \leq b$ and horizontally from $0 \leq x \leq a$.  The flow,  
	
\begin{equation}
	{\bf{V}}(x,y,t) =  \bigg( \frac{\partial \psi}{\partial y}, - \frac{\partial \psi}{\partial x} \bigg) 
\end{equation} 
	
\noindent is defined in terms of a stream function $\psi(x,y)$. The stream function is an exact solution of the biharmonic equation ~$\nabla^2 \nabla^2\psi(x,y) = 0$ defined on the rectangular domain.  The stream function is time-periodic with period $\tau_f$ and is given explicitly by
\begin{equation}\label{stremler} \psi(x,y,t) = \begin{cases}
	U_1C_1f_1(y) \sin \big(\frac{\pi x}{a}\big)  + U_2C_2f_2(y) \sin \big(\frac{2\pi x}{a}\big) , \\~~~~~\quad\quad\mbox{ for } n\tau_f \leq t <(n+1/2)\tau_f,\\
	-U_1C_1f_1(y) \sin \big(\frac{\pi x}{a}\big) + U_2C_2f_2(y) \sin \big(\frac{2\pi x}{a}\big), \\~~~~~\quad\quad\mbox{ for } (n+1/2)\tau_f \leq t <(n+1)\tau_f,
	\end{cases}
\end{equation}  
\noindent where
\begin{equation*}\begin{aligned}
	f_k(y) = &\frac{2\pi y}{a} \cosh\bigg(\frac{k\pi b}{a}\bigg) \sinh\bigg(\frac{k\pi y}{a}\bigg)  \\ & - \frac{2k\pi b}{a} \sinh\bigg(\frac{k\pi b}{a}\bigg) \cosh\bigg(\frac{k\pi y}{a}\bigg), ~~k = 1, 2,
	\end{aligned}
\end{equation*} 
\noindent and
\begin{equation*}
C_k = \frac{a^2}{2k\pi^2 b} \bigg[\frac{a}{2k\pi b}  \sinh \bigg(\frac{2k\pi b}{a}\bigg)  +1\bigg]^{-1} , ~~k = 1, 2.
\end{equation*}

\noindent We follow  Grover et al.\cite{grover2012topological} and assign $U_1 = 9.92786, U_2 = 8.34932, a = 6,$ and $b = 1$.  Fig.~\ref{fig:streamines}a and Fig.~\ref{fig:streamines}b show  streamlines for the two steady flows in Eq.~(\ref{stremler}). Each flow is separately integrable and is asymmetric in $x$, with a large vortex on one side and a smaller vortex on the other. The system alternates between each flow for a half-period $\tau_f/2$. It is this alternating flow that introduces positive topological entropy into the system.

When $\tau_f$ is sufficiently large, $\tau_f \ge \tau_f^* \approx 0.9553$, there exists a period-three orbit, $r_i, ~i = 1, 2, 3$, such that
\begin{equation} \label{per3}
	M(r_1) = r_2, ~~~M(r_2) = r_3, ~~~ M(r_3) = r_1,
\end{equation}
\noindent where $M$ is defined to be the flow map that evolves a point $(x,y)$ forward to the point $(x',y') = M(x,y)$ after a single period $\tau_f$. Fig.~\ref{fig:streamines}c shows the points $r_i$ and their time evolution over one period. In the first half-period, $n\tau_f \leq t <(n+1/2)\tau_f $, the two trajectories on the left swap positions in a clockwise fashion, while in the second half-period, $(n+1/2)\tau_f \leq t <(n+1)\tau_f$,  the two trajectories on the right swap positions in a counterclockwise fashion. Grover et al \cite{grover2012topological} characterize the $r_i$ as a set of three strands braiding around one another in a nontrivial fashion. The presence of this braid guarantees the topological entropy is at least $h_{po3} = 0.9624$,  the topological entropy which Boyland et al.\cite{boyland2000topological, boyland2003topological,boyland1994topological} computed using the Bestvina-Handel train-track algorithm \cite{bestvina1992train}.  We note that this period-three orbit lives within a larger coherent set, a period-three island chain \cite{stremler2011topological} when $\tau_f$ is strictly greater than $\tau_f^*$ .

\subsection{Period-Three Orbit and Convergence in Ensemble Size} \label{subsect:period3}

\begin{figure}[t]
	\centering
	\includegraphics[width=.95\linewidth]{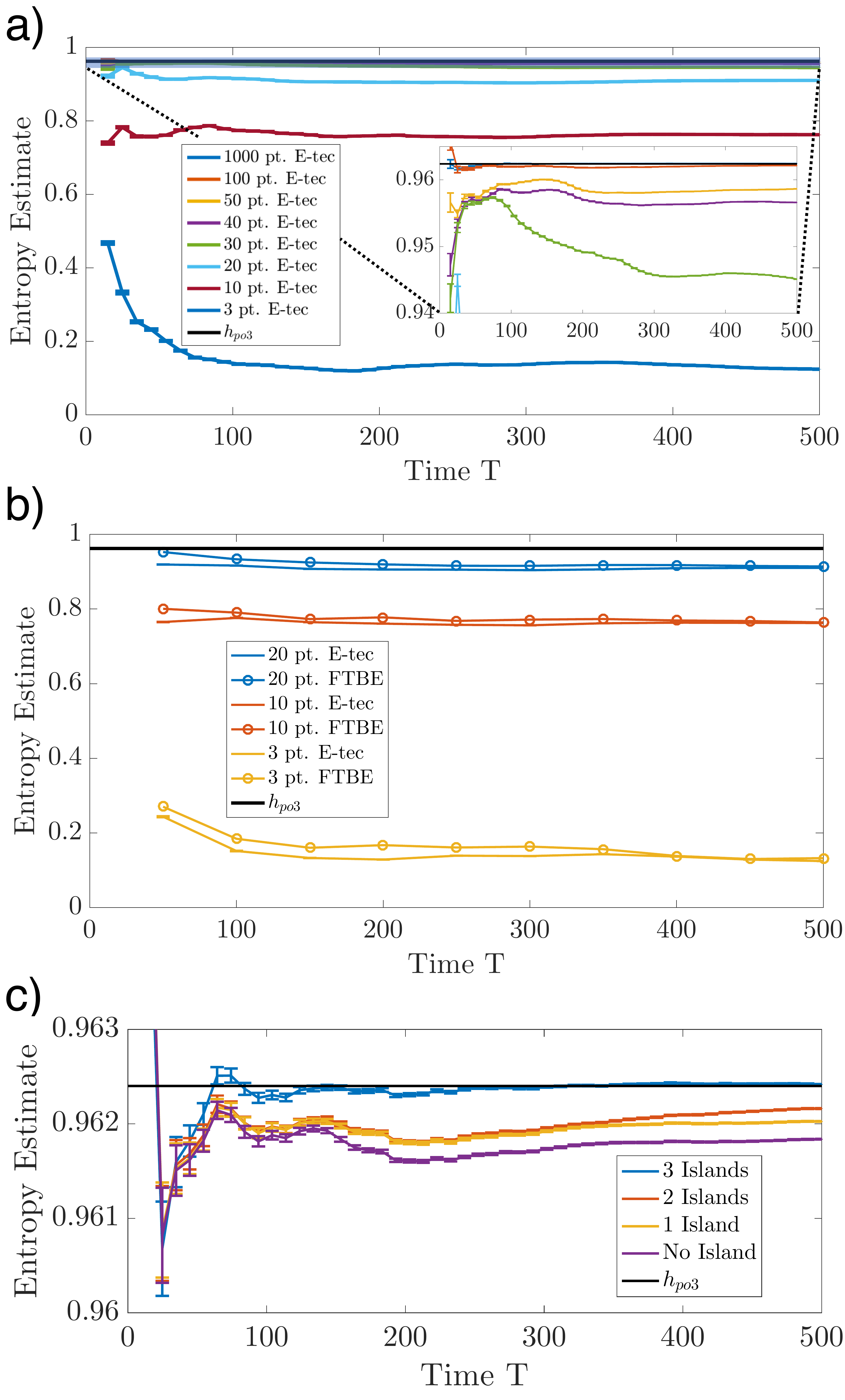}
	
	\caption{\textbf{Convergence of E-Tec in the Length and Number of Trajectories.} ~a) We demonstrate convergence of E-tec to $h_{po3} = 0.9624$ with increasing sample size and trajectory duration. For E-tec, the same initial band is stretched under ensembles of increasing size. All trajectories are sampled from outside the islands in the chaotic lid-driven flow with period-driving parameter $\tau_f = 0.96$. The entropy reported at time $T$ is the fitting slope and 95 percent confidence interval to the log of the total number of edge weights over time $t$ for the range $t \in [5,T]$. ~b) We demonstrate consistency between E-tec and FTBE results (calculated using the freely available Matlab package \texttt{braidlab}\cite{thiffeault2014braidlab}). The same ensembles of trajectories are used for both. To stay consistent with the \texttt{braidlab} calculations, E-tec reports the entropy as the fit from initial time to reported time (or rather, it is the fit from $t \in [0,T]$). ~c) E-tec output using the 100 point ensemble with a single trajectory added into one, two, and three of the periodic islands.}
	
	\label{fig:converge100}
\end{figure}

Here we investigate the convergence of the E-tec algorithm by studying trajectories from the chaotic lid-driven flow with period $\tau_f = 0.96$, where we are guaranteed the existence of a period-three island chain \cite{thiffeault2006topology, grover2012topological, gouillart2006topological}. As illustrated in Fig.~\ref{fig:etecper3}d, no trajectory starting in an island leaves the island, and no trajectories enter. These islands braid around one another as they swap places in the same fashion depicted in Fig.~\ref{fig:streamines}c. In the analysis of Sect.~\ref{sect:stremler}, each trajectory is sampled with time step $\Delta t = 10^{-2}$ between points. This choice of $\Delta t$ will be shown to be sufficient in Sect.~\ref{sect:stepsize}.

First, we run E-tec on a set of three trajectories with the initial condition for each trajectory chosen in a different period-three island (Fig.~\ref{fig:etecper3}a). We place an initial band around the right two points and observe exponentially growing band weights (Fig.~\ref{fig:etecper3}b). At $T = 15$ our estimate for the topological entropy is within $0.1\%$ of the topological entropy guaranteed by the braid (Fig.~\ref{fig:etecper3}c). 

Next, we run E-tec on a set of 75 trajectories consisting of the 3 previously selected trajectories along with 72 randomly chosen ones. We calculate topological entropy by considering the time evolution of the same initial band (Fig.~\ref{fig:etecper3}b). While the dynamics appear far more complicated than in Fig.~\ref{fig:etecper3}a, our estimate of topological entropy is within fitting error to $h_{po3} = 0.9624$ (Fig.~\ref{fig:etecper3}c).  Our results demonstrate that the periodic islands, and their braiding, are what drives most of the system entropy \cite{li1975period, sharkovsky1964coexistence}. Furthermore, this demonstrates that for certain systems, topological approaches such as E-tec (as well as braiding approaches) are capable of producing accurate estimates of topological entropy with only a small set of carefully chosen trajectories.
		
Although the coherent sets for our example were straightforward to locate, for other examples and practical applications, coherent sets may be harder to identify. As such, there is no guarantee trajectories from coherent sets, whose dynamics might be governing the topological entropy of the system, will be sampled appropriately. To investigate how E-tec would perform under conditions like this, we examine our ability to accurately recover the topological entropy when randomly sampling initial conditions uniformly in space, but removing any point chosen in the period-three islands. E-tec was run on increasingly larger but nested sets of such trajectories. That is, the points chosen in the 20 trajectory analysis contain all of the points in the 10 trajectory analysis, and so forth. As shown in Fig.~\ref{fig:converge100}a, E-tec converges rather quickly in the number of points to the topological entropy lower bound guaranteed by the period-three islands. Estimates may fluctuate based on the interval used to fit, especially when fewer trajectories are used. In Fig.~\ref{fig:converge100}, we see apparent oscillatory behavior, though we expect these to dampen at longer times and for results to converge if taken to infinite time. We note that in the above figure that E-tec does not require many long trajectories to compute a reasonable approximation to the topological entropy.

Finally, in Fig.~\ref{fig:converge100}b, we investigate the E-tec convergence using the 100 point ensemble in Fig.~\ref{fig:converge100}a by adding additional points in each of the three islands. E-tec performs increasingly better as the island points are added. The result with no island points, given in Fig.~\ref{fig:converge100}a, is then taken as a worst-case scenario. This assures our confidence in E-tec results as ensemble sizes are increased in Sect.~\ref{subsect:entropyRange}.

\subsection{Topological Entropy for Range of Period Driving Parameter $\tau_f$} \label{subsect:entropyRange}

\begin{figure}[t]
	\centering
	\includegraphics[width=1\linewidth]{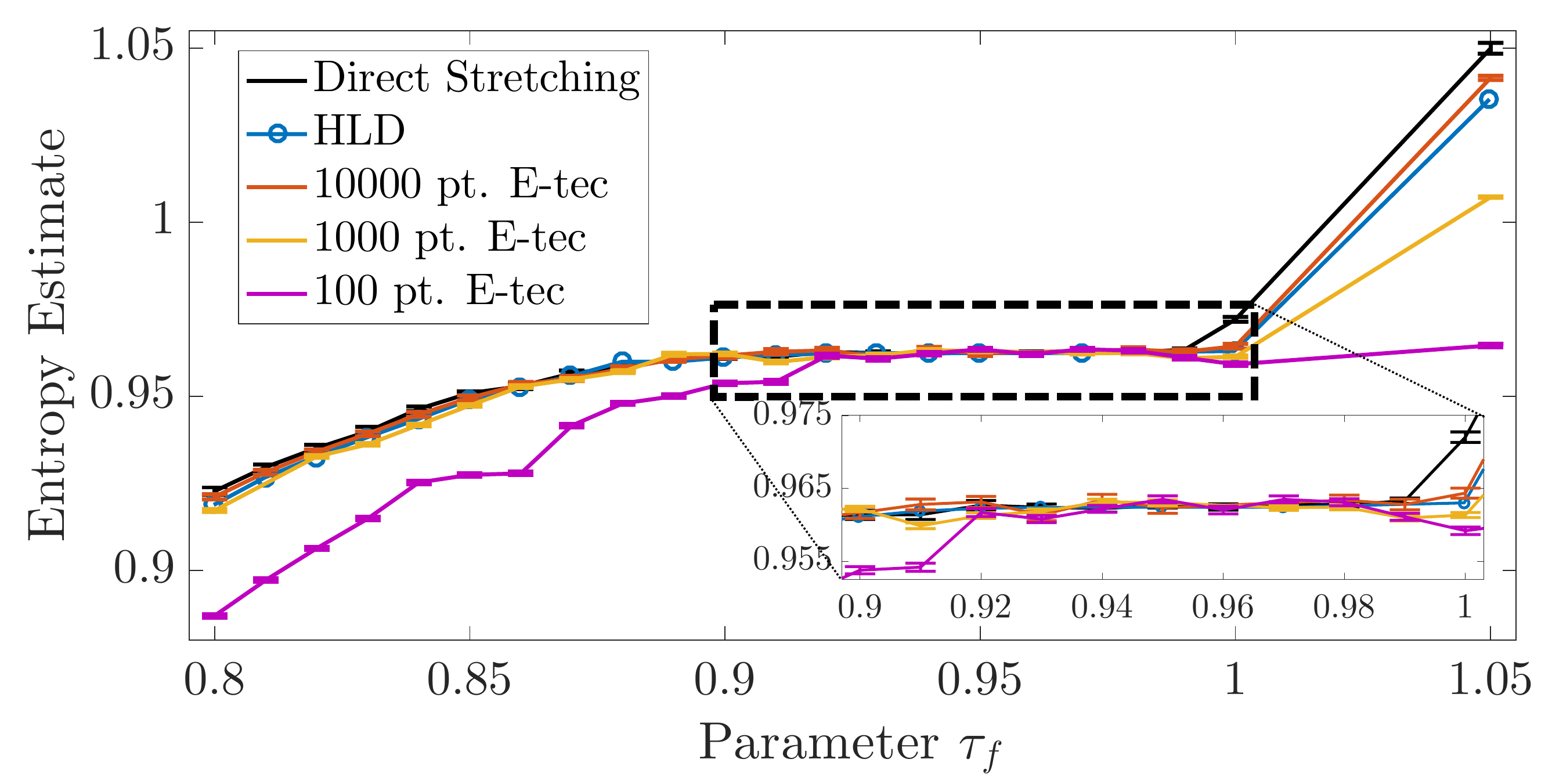}
	
	\caption{\textbf{Verification of E-tec for Increasing $\tau_f$.} E-tec topological entropy results over a range of $\tau_f$ values using increasing ensemble sizes.  We compare to the estimate of topological entropy from directly stretching a material line \cite{sattari2016using} and through another topological technique, homotopic lobe dynamics \cite{mitchell2009topology, mitchell2012partitioning}. }
	
	\label{fig:entropyconvergence}
\end{figure}

With confidence in E-tec's ability to characterize topological entropy when $\tau_f = 0.96$, we  next explore how the topological entropy changes as $\tau_f$ varies.  As mentioned previously, the period-three orbit is born at $\tau_f^* \approx 0.9553$ and persists for larger values. Thus, entropy for values $\tau_f < \tau_f^*$ will be bounded above by the braiding entropy of $h_{po3} = 0.9624$, while $h_{po3}$ remains a lower bound for $\tau_f > \tau_f^*$. In all cases, the same initial band is chosen and evolved forward.

As shown in Fig.~\ref{fig:entropyconvergence}, our estimate of topological entropy using E-tec is within error of the direct calculation of material-line stretching when $0.85 \leq \tau_f \leq 0.98$ and the number of data points is at least 1000. For $\tau_f < \tau_f^*$, there are no known island chains that drive the complexity. Despite this, E-tec performs well here, as shown in Fig.~\ref{fig:entropyconvergence}. For low values of $\tau_f$, when $\tau_f < 0.85$, E-tec produces an estimate slightly less than that of direct stretching but consistent with the value produced by HLD. But E-tec's discrepancy becomes smaller with increasing numbers of samples.  For high values of $\tau_f$, when $\tau_f > 0.98$, both E-tec and HLD produce lower estimates for topological entropy than the calculated direct stretching value. We note that E-tec with 1000 trajectories still produces estimates consistent with HLD, and with 10,000 trajectories E-tec exceeds the HLD estimate but is still below the direct material-line stretching. 

\begin{figure}[t]
	\centering
	\includegraphics[width=1\linewidth]{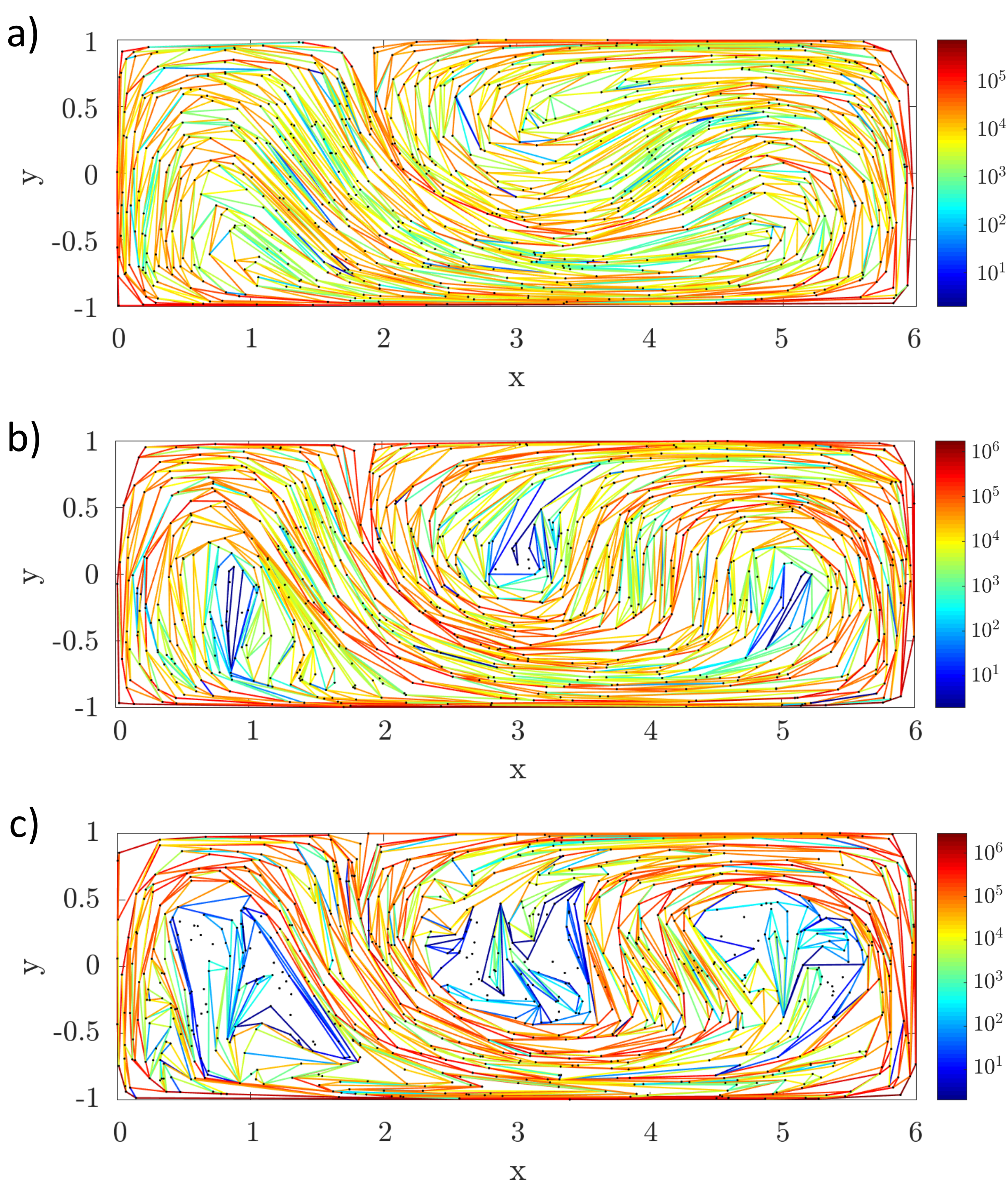}
	\caption{\textbf{Stretched Band Visualization.} E-tec band stretching due to flow advection for period-driving parameters ~a) $\tau_f = 0.80$, ~b) $\tau_f = 0.96$, and ~c)  $\tau_f = 1.05$. Colorbar corresponds to edge weights. All bands are stretched by ensembles of 1000 uniformly distributed trajectories.}
	\label{fig:bandcomparison}
\end{figure}

To more clearly see what drives the increase in entropy for high values of $\tau_f$, we show the band stretched by E-tec for three different values of $\tau_f$ each computed from a set of 1000 independently chosen trajectories (see Fig.~\ref{fig:bandcomparison}). Exponential stretching and folding is present in all tested parameter values, though Fig.~\ref{fig:bandcomparison} shows the band is stretched in a more complex fashion at higher $\tau_f$ values. Here, additional island chains emerge\cite{stremler2011topological} resulting in secondary folding \cite{tumasz2013estimating} that seems less "smooth." This secondary folding results in kinks near the islands that propagate forward, which in turn are further stretched under the dynamics.  These small areas with kinks give significant contribution to the topological entropy, but because the entropy estimates (Fig.~\ref{fig:entropyconvergence}) were generated from uniformly random samples, these highly-kinked regions may remain undersampled. As such, a good portion of the stretching may remain undetected by E-tec in Fig.~\ref{fig:bandcomparison}c.

\section{E-tec Robustness} \label{sect:discussion}

In this section, we investigate the robustness of E-tec's results. More specifically, we examine how E-tec's ability to correctly estimate topological entropy is impacted by the choice of initial band and the time-step associated with trajectories. Finally, we discuss how the E-tec algorithm's run-time scales with the duration and number of sampled trajectories. 

\subsection{Robustness to Choice of Initial Band} \label{sect:initialband}

\begin{figure}[t]
	\centering
	\vspace{5pt}
	\includegraphics[width=1\linewidth]{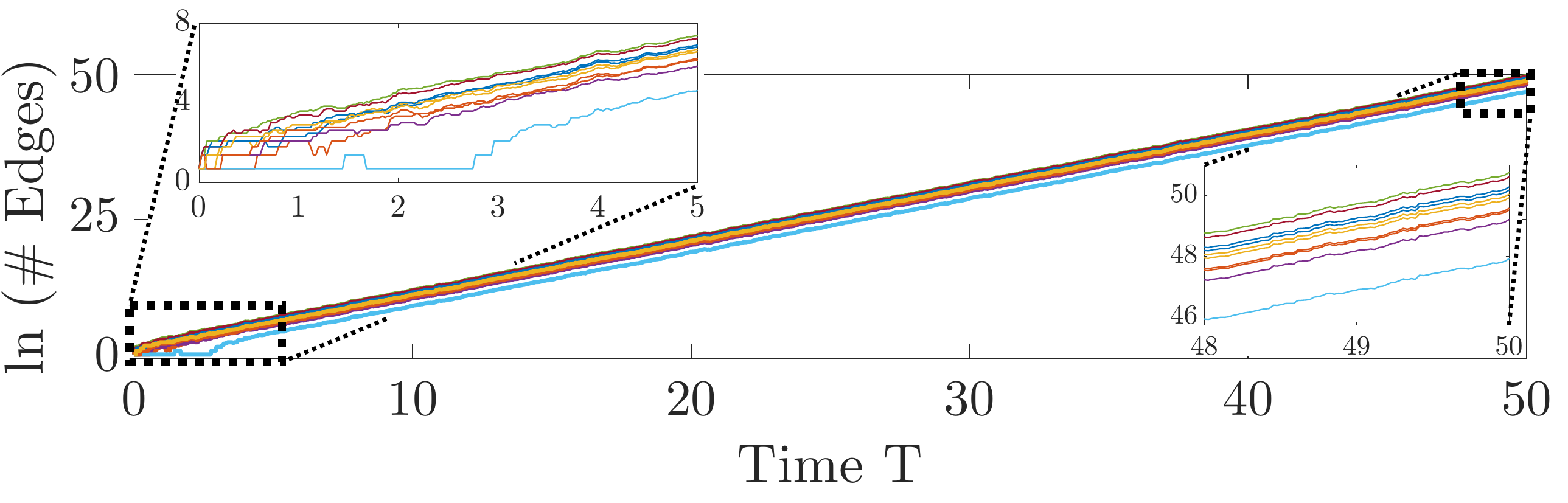}
	
	\caption{\textbf{Initial Bands.} E-tec output is the logarithm of the sum of edge weights as a function of time. E-tec's estimate of topological entropy is the best-fit linear slope through this data. Here we show 10 different outputs from E-tec for the same set of 100 trajectories. In each case, we chose a different pair of points around which to stretch our band. Despite some initial differences in the increase in edge weights due to initial adjacent points staying close to one another (\emph{left} inset), eventually all the bands grow at similar rates (\emph{right} inset). When we fit the exponential growth rate, starting at $T = 5$, we find the values for each of the 10 bands agree within 5 decimal places and average out to $0.9617$.}

	\label{fig:initialbands}
\end{figure}

We make the following conjecture: if all trajectories reside in the same ergodic component then the choice of initial band does not affect the topological entropy computed by E-tec as long as the trajectories are sufficiently long.  Figure~\ref{fig:initialbands} supports this conjecture. All initial bands eventually become stretched at the same rate despite some differences at early times. Adjacent points may remain close for some time, though the chaotic nature of the flow causes nearby trajectories to eventually diverge, thereby making the band's deformation inevitable. Thus, as long as it is possible to obtain sufficiently long trajectories within a single ergodic component, E-tec's topological entropy calculation appears to be invariant to the choice of initial band. 

Some chaotic flows have more than one ergodic component, or a mixture of ergodic and non-ergodic regions. This is true of the model flow in Fig.~\ref{fig:etecper3}d. In such systems, the choice of initial band will impact the topological entropy estimate. For example, a band placed entirely in one of the test flow's period-three islands (Fig.~\ref{fig:etecper3}d) will undergo no significant stretching under the flow and thus yield zero topological entropy. 

In practice, to make sure all ergodic components are sampled, it is prudent to check that the final band stretches around nearly all of the data points. Alternatively, one could sample many initial bands taking the maximum growth rate of all sampled bands as the best estimate of the entropy \cite{thiffeault2005measuring}. E-tec is fast enough to run multiple bands, each with a a different initial triangulation constrained to the initial band choice, in ensembles of fewer than $10^6$ trajectories in a reasonable time. An alternative approach to choosing a single initial band is to evolve a "web" of initial bands that covers the entire initial triangulation. This guarantees that all ergodic components sampled by the data will be included. As opposed to the initial triangulation being constrained to the choice of initial band, the initial "web" is constrained to the edges of the choice of initial triangulation.

\subsection{Algorithm Scaling and FTBE Comparison} \label{sect:scaling}

The computational runtime of E-tec is linearly proportional to the \emph{duration} of the trajectories. This is because the number of edges tracked by E-tec is constant, and it is only the values of the weights that grow exponentially in time. This scaling is the same as the FTBE calculation and stands in contrast to algorithms that precisely evolve a material-curve forward, which requires inserting exponentially more points to maintain sufficient point density~\cite{you1991calculating}. 

\begin{figure}[t]
	\centering
	\includegraphics[width=1\linewidth]{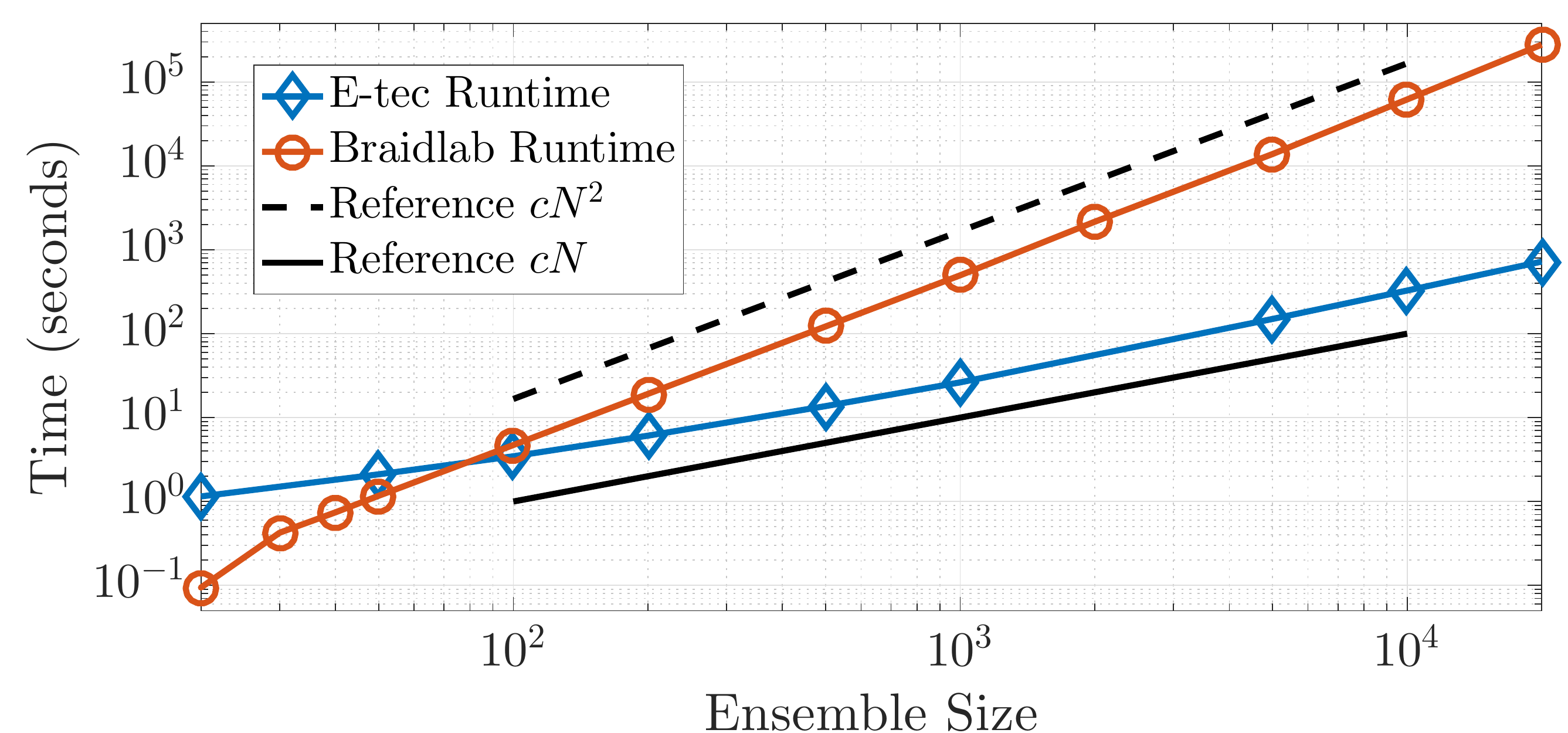}
	
	\caption{\textbf{E-tec Runtimes.} Runtime comparison of E-tec and \texttt{braidlab}, a freely available Matlab package implementing the FTBE calculation. Both used the same trajectories from the chaotic model flow for $\tau_f = 0.96$. All computations were completed using a 2.8 GHz Intel Core i7 processor.}
	\label{fig:runtime}
\end{figure}

One advancement we have made over the FTBE calculation is the run-time scaling with respect to the number of trajectories used (see Fig.~\ref{fig:runtime}). The FTBE calculation scales quadratically in the number of trajectories $N$ due to the braid approach requiring $N^2$ algebraic generators per unit time step \cite{budivsic2015finite}. Overall, E-tec runtime scales as $\mathcal{O}(N + N^k \log{}N)$, where $k$ is the collapse event rate scaling factor. In general, the value of $k$ largely depends on the complexity of the flow being studied. For the chaotic, lid-driven cavity flow trajectories, we find $k \approx 1.05$, though we find values as low as $k = 1/3$ for trajectories with highly correlated movement and as high as $k = 3/2$ for random trajectories. We refer the reader to Appendix~\ref{sect:appendix} for more details. As a practical matter, the E-tec runtime for small to moderate ensembles (roughly up to 5,000 trajectories) is dominated by the linear behavior in Fig.~\ref{fig:runtime}. 


One illustrative example highlighting the runtime difference between the two algorithms is rigid rotational flow. While an admittedly special case, there would be no new collapse events (except for ones associated with the boundary) making E-tec very fast, whereas the number of braid generators needed would be proportional to $N^2$. However, one advantage the braid approach has over E-tec is that once the braid is extracted from the trajectory data, it may be applied to any initial band.  E-tec only propagates a single curve forward. However, for topological entropy calculations, a single sufficiently long curve is typically sufficient (as evidenced in Fig.~\ref{fig:initialbands}). 

\subsection{Robustness to Step Size $\Delta t$}\label{sect:stepsize}
	
\begin{figure}[t]
	\centering
	\includegraphics[width=1\linewidth]{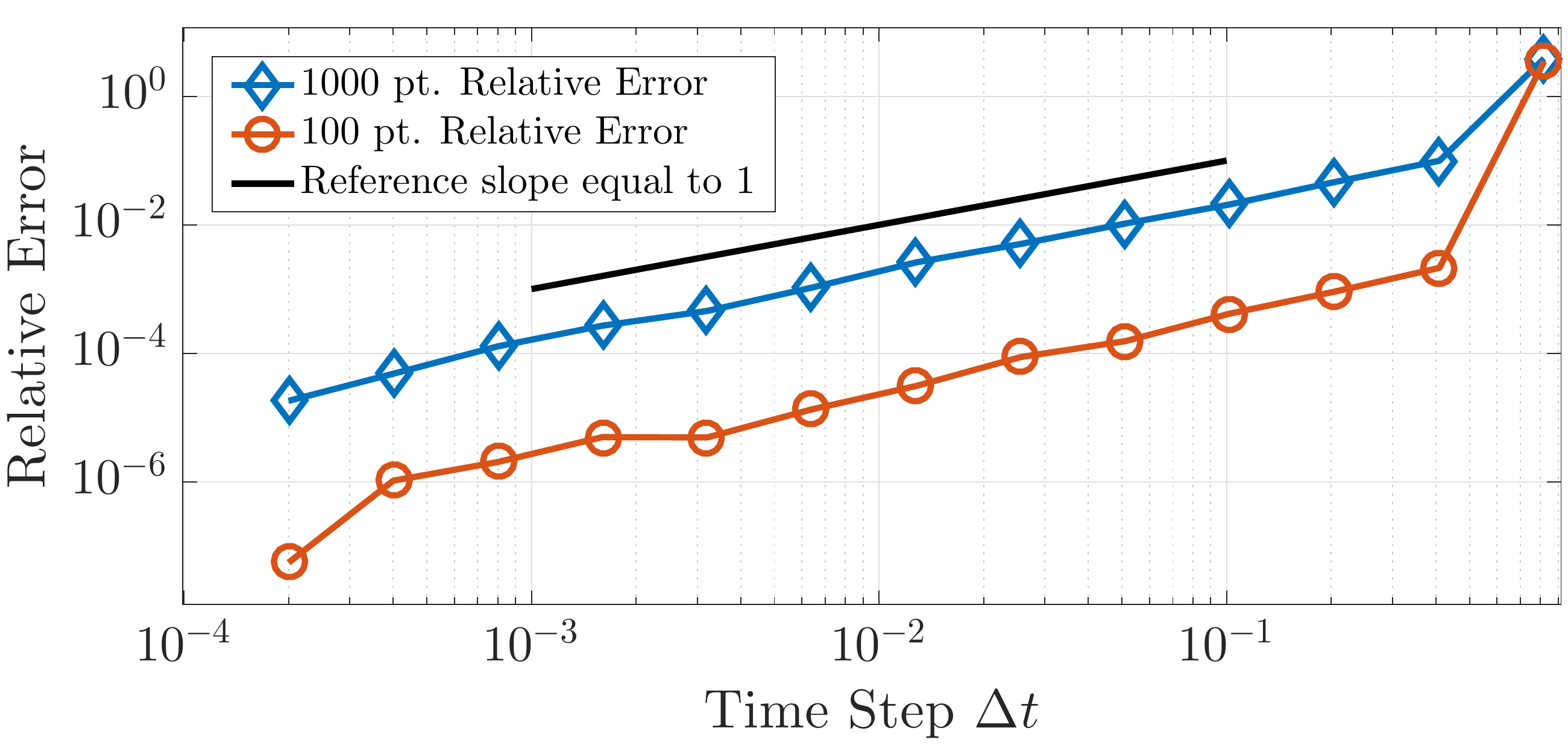}
	\caption{\textbf{Relative Error as a Function of Step Size.} The effect of time step $\Delta t$ on the relative error in the topological entropy calculation with respect to the reference time step $\Delta t^* = 10^{-4}$.  Graph displays calculations done on two separate ensembles of size 1000 and 100.}
	\label{fig:decimate}
\end{figure}

Because E-tec is based on the computational analysis of evolving trajectories, it is necessary to consider discretized time. We next investigate how the trajectory time step $\Delta t$ affects the entropy calculation and show that E-tec returns trustworthy results even when poorly resolved trajectories are used as input. We use two ensembles of trajectories (of sizes 100 and 1000) sampled at a fine scale using the same reference time step of $\Delta t^*= 10^{-4}$ to generate two reference topological entropies $h_t^*$. We then vary the time step $\Delta t$ (keeping the trajectories the same) and compute both ensembles' corresponding $h_t$. The effect of time step $\Delta t$ is quantified by computing the relative error \begin{equation} \label{eq:relerror}
	\bigg|1 - \frac{h_t}{h_t^*} \bigg|,
	\end{equation} 
	
\noindent which is plotted in Fig.~\ref{fig:decimate}. The data shows the relative error grows linearly with the time step $\Delta t$.   As the trajectory information is input into E-tec using larger step sizes, we detect more events between steps. E-tec detects events individually for all values of $\Delta t$, but the order in which these events are detected is potentially different as $\Delta t$ increases, due to the differences in the interpolation of trajectories. In fact, undersampled trajectory data may lead to entirely different events. This explains the larger relative errors for the 1000 trajectory ensemble; at higher point densities, there are simply more events that E-tec must resolve, resulting in more erroneous and misordered event detections. Despite this, Fig.~\ref{fig:decimate} shows that the E-tec error due to step size is still relatively small.  It is comparable to (or smaller than) the error due to other sources, such as trajectory length and ensemble size (Fig.~\ref{fig:converge100}a), for $\Delta t < 10^{-2}$, at least for smaller ensemble sizes.

\section{Conclusion} \label{sect:conclusion}
We introduced the Ensemble-based Topological Entropy Calculation (E-tec), an algorithm that computes topological entropy in a planar flow from an ensemble of system trajectories. We verified E-tec's convergence to the correct topological entropy with increasing numbers of trajectories on a highly chaotic, lid-driven cavity flow. E-tec's performance was shown to be robust with respect to the choice of initial band, as well as changes in the time sampling interval $(\Delta t)$.  Notably, we have shown that E-tec's runtime scales as $\mathcal{O}(N^k \log N)$, where $1/3 \leq k \leq 3/2$ and $N$ is the number of trajectories in the ensemble. 

Our work suggests several further directions for the analysis of trajectories with E-tec, which we intend to explore in future studies. First, we shall seek to extend E-tec to three dimensions and higher. Braiding theory, the basis for FTBE calculations, cannot be readily generalized to higher dimensions\cite{allshouse2015lagrangian}. The computational geometry framework in which E-tec is based might perhaps be more naturally extended \cite{lefranc2013reflections, lefranc2006alternative, lefranc2008topological}. Instead of a rubber band in a planar flow, we would consider a two-dimensional rubber sheet stretched around a collection of points in a three-dimensional flow. A 3D triangulation may still be used to track point-face or edge-edge collisions, and the rubber sheet may be chosen as one of the faces in the initial triangulation. As the points evolve in time, they carry the sheet along with them, stretching and folding it so that its growth reflects the flow complexity. Though there clearly remain some significant challenges to executing this generalization to three dimensions, we anticipate a host of interesting theoretical opportunities that this route may provide. Finally, by tracking all the trajectories in concert, we believe E-tec's algorithm may be naturally adapted towards identifying and tracking coherent sets and other emergent structures. 




\section*{Acknowledgments}

This work was supported in part by the US DOD, ARO grant W911NF-14-1-0359 under subcontract C00045065-4. ER also received support from the National Science Foundation under grant number DMS-1331109.

\appendix
\section{ E-tec runtime scaling with number of points}  \label{sect:appendix}

The main bottleneck in the computational complexity of E-tec comes from the creation and maintenance of a time-sorted collapse-event list at each time-step.  Since every core and outer triangle, of which there are $\mathcal{O}(N)$, is checked for collapse in this process, E-tec will scale no better than linear in $N$.  Sorting is a worst-case and average-case $\mathcal{O}(n \log{}n)$ process, for $n$ items to sort.  Assuming that the number of collapse events per unit time scales as  $\mathcal{O}(N^k)$ for some $k$, the sorting bottleneck implies an E-tec scaling of $\mathcal{O}(N^k \log{}N)$.  A similar scaling comes from the maintenance of this event list.  During the handling of a collapse event, core and outer triangles may be created, modified, or deleted.  Importantly, this process is local, and the time for handling one event does not change with an increasing number of points.  However, these amendments to the triangulation necessitate adding or removing events from the time-sorted event list.  This is achieved with a binary search, which is an $\mathcal{O}(\log{}n)$ routine for a list length of $n$.  Given a list length that scales with the number of collapse events per unit time, this constitutes a second avenue for the $\mathcal{O}(N^k \log{}N)$ scaling.  Overall, the E-tec runtime scales as $\mathcal{O}(N + N^k \log{}N)$, where $k$ is determined by the collapse event rate scaling.

\begin{figure}[t]
	\includegraphics[width = 0.5 \textwidth]{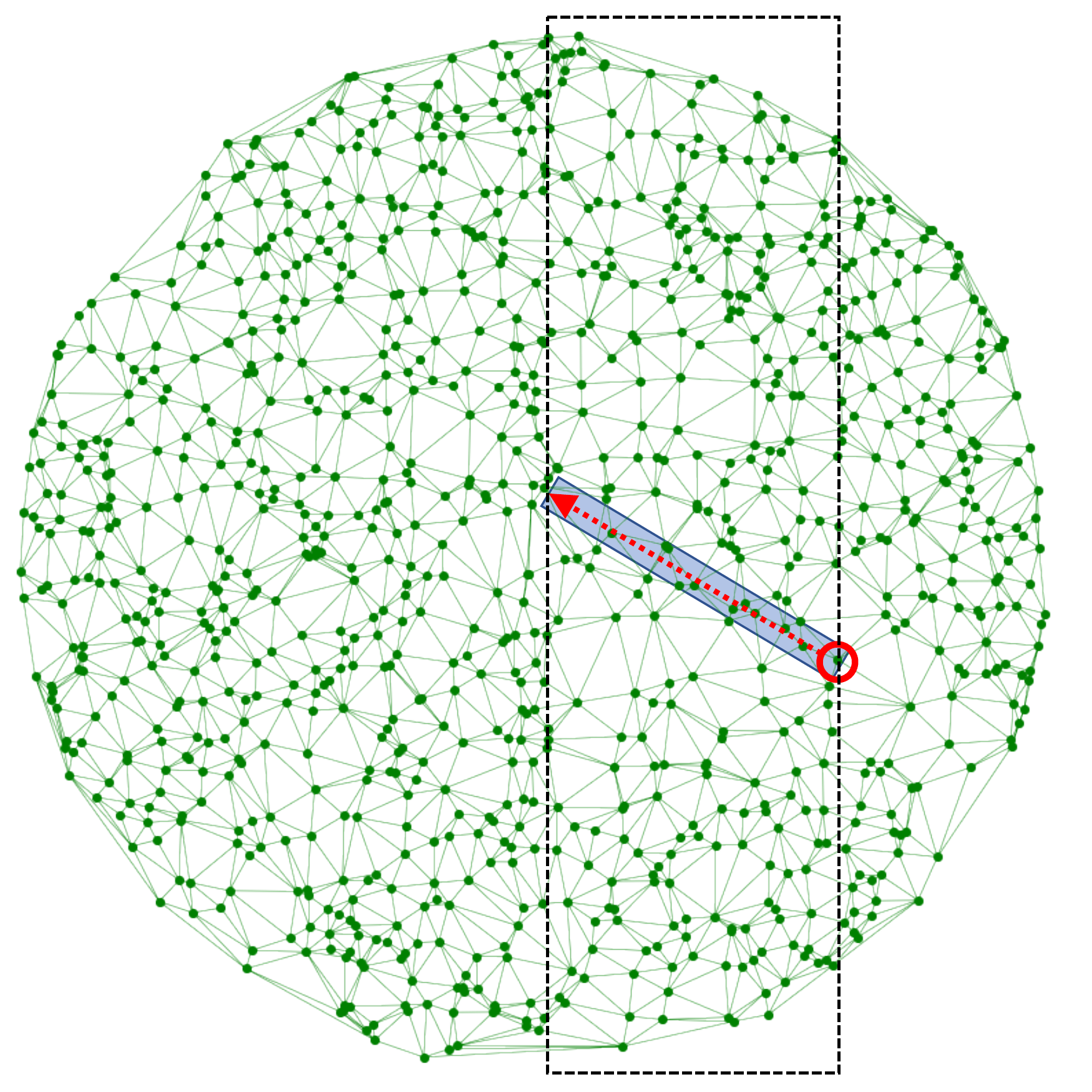}
	\footnotesize
	\caption{Schematic of the movement of a single point (circled in red) against a background of stationary points.  The number of E-tec collapse events is proportional to the number of average core triangles that would have an equivalent area to that of the diagonal blue rectangle.  For FTBE calculations, the number of braid generators created by this same process is equal to the number of points in the larger black rectangle.}
	\normalsize
	\label{TriangulationFig}
\end{figure}

The scaling of the collapse event rate depends heavily on the type of flow that produced the trajectory data.  If there is no correlation between the velocities of neighboring points, then it can be as high as $k=3/2$.  If they are highly correlated (e.g. rigid rotation), then it can be as low as $k=1/3$.  For most flows, $k \lesssim 1$, with $k$ generally increasing for more complex flows.  

To justify the worst-case scaling of $k = 3/2$, consider the movement of a single point through a fixed length and against a background of stationary points, as depicted in Fig.~\ref{TriangulationFig}.  The number of collapse events produced by this motion will be proportional to the number of core triangles in the path of the moving point.  Given that the average area of a core triangle scales as $\mathcal{O}(N^{-1})$, a characteristic triangle length goes as $\mathcal{O}(N^{-1/2})$. Therefore, the number of characteristic lengths in the particle's path, and from this the number of collapse events, scales as $\mathcal{O}(N^{1/2})$.  Moving to the general case where every point is in motion, we could say that each of the $N$ points "sees" $\mathcal{O}(N^{1/2})$ triangles in its way, and therefore the overall scaling for collapse events would be $\mathcal{O}(N^{3/2})$.  For comparison with FTBE calculations, the same one-point motion produces a number of braid generators equal to the number of points in the black rectangle of Fig.~\ref{TriangulationFig}.  Since this scales as $\mathcal{O}(N)$, the general case where every point is in motion produces a braid generator production rate that scales as $\mathcal{O}(N^2)$.

\begin{figure}[t]
	\includegraphics[width = 0.5 \textwidth]{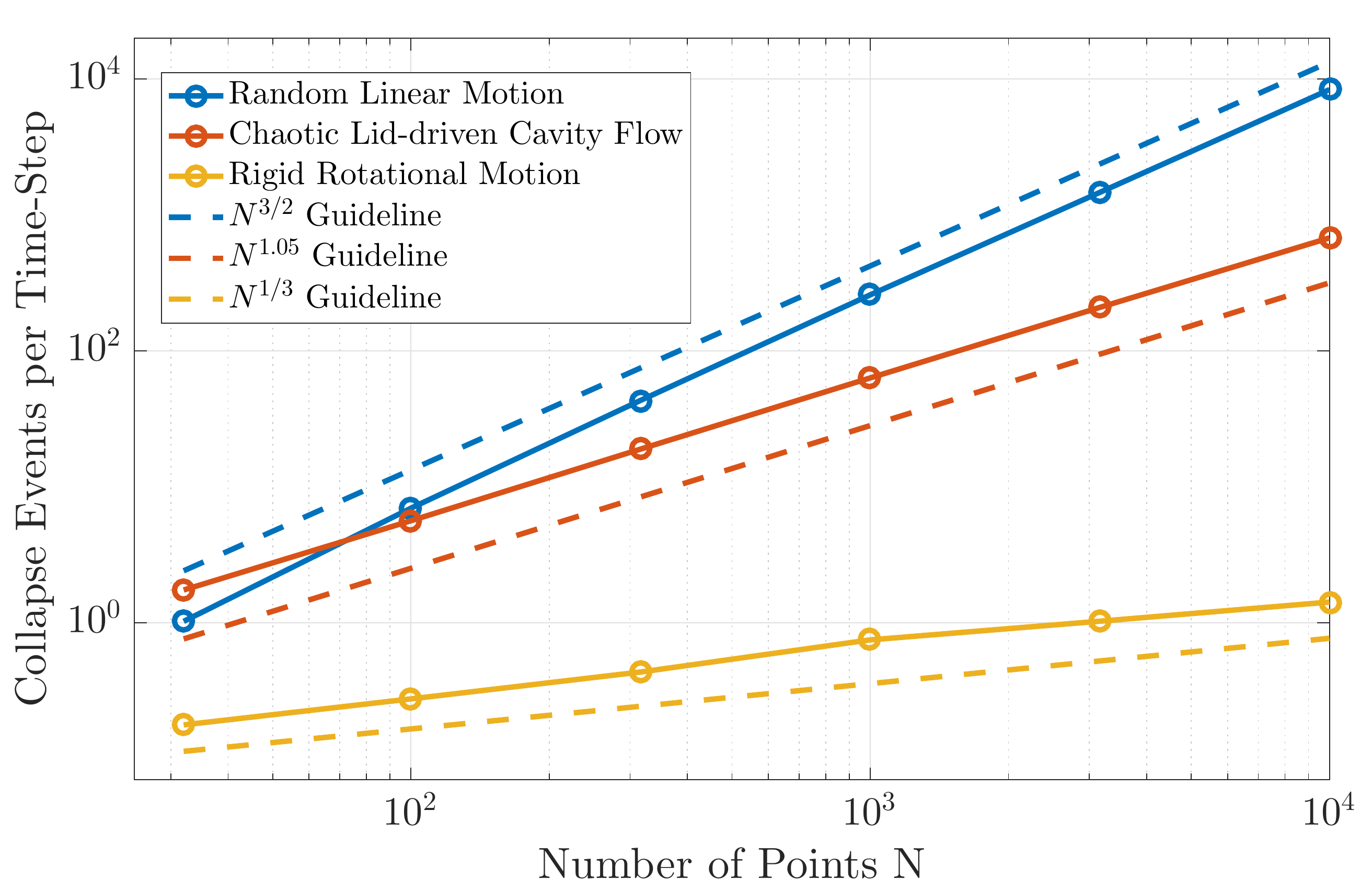}
	\caption{The collapse event rate scaling for three numerical examples:  random motion, chaotic lid-driven cavity flow (for $\tau = 0.96$), and rigid body rotation.}
	\label{3Examples}
\end{figure}

The $k = 3/2$ scaling is also borne out in a numerical experiment, see Fig.~\ref{3Examples}.  Here we track the collapse event rate for $N$ trajectories, whose initial and final positions are chosen randomly within a fixed square, and whose intermediate positions are given by linear interpolation.

However, in most cases of interest the trajectory motion is generated by or sampled from an underlying flow, and there will be substantial correlations between the movement of nearby points.  Points advected together can significantly suppress the collapse event rate scaling.  At the other extreme, consider the case of points undergoing rigid-body rotation.  None of the triangles in the bulk will collapse, and the only contribution to the collapse event rate comes from core triangles associated with the fixed bounding auxiliary points (stationary points that are added upon initialization which help us avoid triangulation update issues at the boundary edges of the triangulation).  This numerical example, see Fig.~\ref{3Examples}, gives a scaling value of $k \approx 1/3$, likely the most favorable scaling we can expect from a non-trivial flow.  General flows will fall between these two extremes.  Our example of a chaotic lid-driven cavity flow (see Fig.~\ref{3Examples}), with $\tau = 0.96$, gives a scaling value of $k = 1.05$.  We have also simulated the collapse event rate scaling for linear shear flow, $k \approx 0.66$ and an irrotational (Rankine) vortex, $k \approx 0.77$.

Overall, we can expect the E-tec runtime to scale as 	$\mathcal{O}(N + N^k \log{}N)$, with $1/3 \leq k \leq 3/2$, and typical flows resulting in $k \lesssim 1$.  This favorable computational complexity, compared to $\mathcal{O}(N^2)$ for the FTBE, comes from two sources.  First, collapse events are produced locally, whereas braid generators encode more global information.  Second, the correlated motion of neighboring points further reduces the scaling for trajectories derived from general flows.

\bibliographystyle{aipnum4-1}
\bibliography{E-tec_References}

 \end{document}